\documentclass{article}
\usepackage[utf8]{inputenc}

\usepackage{alphabeta}
\usepackage{amssymb}
\usepackage[T1]{fontenc}    
\usepackage{hyperref}       
\usepackage{url}            
\usepackage{booktabs}       
\usepackage{amsfonts}       
\usepackage{nicefrac}       
\usepackage{microtype}      
\usepackage{amsmath}
\usepackage{amssymb}
\usepackage{amsmath, amsthm}
\usepackage{graphicx}
\usepackage{color}
\usepackage[colorinlistoftodos]{todonotes}
\usepackage{datetime}
\theoremstyle{definition}

\title{IoT Integration Protocol for Enhanced
Hospital Care}

\author{Ellie Zontou\\
  Dept. of Computer Engineering and Informatics\\
  University of Patras\\
  Patras, 26504, Greece \\
  \texttt{ezontou@ceid.upatras.gr} \\
\and
 Antonia Kyprioti \\
  Dept. of Medicine\\
  University of Thessaly\\
  Larisa, 41334, Greece \\
  \texttt{akyprioti@uth.gr}}

\begin{document}
\maketitle

\begin{abstract}
This paper introduces the "IoT Integration Protocol for Enhanced Hospital Care", a comprehensive framework designed to leverage Internet of Things (IoT) technology to enhance patient care, improve operational efficiency, and ensure data security in hospital settings. With the growing emphasis on utilizing advanced technologies in healthcare, this protocol aims to harness the potential of IoT devices to optimize patient monitoring, enable remote care, and support clinical decision-making. By integrating IoT seamlessly into nursing workflows and patient care plans, hospitals can achieve higher levels of patient-centric care and real-time data insights, leading to better treatment outcomes and resource allocation. This paper outlines the protocol's objectives, key components, and expected benefits, while emphasizing the importance of ethical considerations and ongoing evaluation to ensure successful implementation.
 
\end{abstract}

\small{\it IoT Integration Protocol
, Enhanced Hospital Care
, Internet of Things (IoT) technology
, Patient care optimization
, Remote care
}

\section{Introduction}

Advancements in technology have been reshaping the landscape of healthcare, revolutionizing patient care, and transforming operational efficiencies within hospital settings. Among these technological innovations, the Internet of Things (IoT) stands as a pivotal force, offering a promising avenue for enhancing hospital care delivery. This paper introduces the "IoT Integration Protocol for Enhanced Hospital Care", an innovative framework tailored to harness the potential of IoT technology within healthcare contexts.

The healthcare sector is undergoing a significant paradigm shift, influenced by the integration of cutting-edge technologies. This transformative wave is propelled by the increasing demand for precision medicine, patient-centered care, and streamlined healthcare operations. Technologies such as IoT have emerged as crucial enablers, offering unprecedented opportunities to augment patient monitoring, facilitate remote care services, and empower healthcare professionals with real-time data insights.

 The significance of IoT in Hospital Settings
IoT devices, characterized by their interconnectedness and data-sharing capabilities, hold immense promise in the context of hospital care.  For instance, wearable biosensors, as discussed in \cite{Xiao}, can track physiomes and activity, providing valuable health-related information that contributes to informed decision-making and personalized care. 
These devices encompass a wide array of interconnected sensors, actuators, and wearable devices, collectively capable of capturing, transmitting, and analyzing patient data in real-time. In addition to wearable biosensors, IoT-based smart healthcare systems have been shown to improve the quality of care through enhanced data analytics and real-time monitoring \cite{Alekya}.  This wealth of information provides a foundation for informed decision-making, personalized care, and proactive interventions, ultimately leading to improved patient outcomes.

The primary focus of this paper is to introduce and elucidate the "IoT Integration Protocol for Enhanced Hospital Care". Within this framework, the paper aims to dissect the key components, delineate the expected benefits, and emphasize the crucial role of ethical considerations and continuous evaluation in successful implementation.
Each subsequent chapter delves deeper into specific aspects of the IoT Integration Protocol. Chapter 2 elucidates the fundamental objectives and importance of IoT within hospital settings. Chapter 3 examines the key components and strategies involved in integrating IoT devices into nursing workflows and patient care plans. Chapter 4 explores challenges and implementation strategies. Chapter 5 assesses the impact and effectiveness of the IoT Integration Protocol. Chapter 6 provides future directions and recommendations for healthcare IoT integration, ensuring a comprehensive view of its potential in enhancing hospital care.

\section{The Significance of IoT in Hospital Settings}

The integration of Internet of Things (IoT) technology in hospital settings ushers in a new era of healthcare delivery, marked by continuous monitoring, remote care services, operational efficiency, and data-driven decision-making. This introduction offers a comprehensive view of the importance of IoT in hospitals, emphasizing its transformative influence on patient care, operational workflows, and overall healthcare outcomes.

\begin{itemize}

\item{Transforming Patient Monitoring and Care Delivery}\\
In hospital settings, the ability to monitor patients continuously and in real-time is crucial for delivering high-quality care. Traditionally, patient monitoring has been limited to periodic checks by healthcare staff or the use of stationary monitoring equipment. However, IoT technology enables a paradigm shift by allowing for continuous, remote monitoring of patients using interconnected devices \cite{Xiao}.

IoT devices such as wearable sensors, smart beds, and connected medical equipment can collect a wide range of patient data, including vital signs, movement patterns, and medication adherence. This real-time data can provide healthcare providers with valuable insights into patient health status, allowing for early detection of complications, timely interventions, and personalized treatment plans (Chen et al., 2018). By leveraging IoT technology, hospitals can improve patient outcomes, reduce the risk of adverse events, and enhance the overall quality of care delivery.

\item{Enabling Remote Care Services and Telemedicine}\\
The adoption of IoT technology in hospitals extends beyond the confines of traditional healthcare settings, enabling the delivery of remote care services and telemedicine solutions. Remote patient monitoring (RPM) solutions, powered by IoT devices and telehealth platforms, allow patients to receive care from the comfort of their homes while remaining connected to their healthcare providers.

For patients with chronic conditions, remote monitoring solutions offer the opportunity for proactive disease management and early intervention, leading to better health outcomes and reduced healthcare costs.  This is exemplified in studies such as \cite{Bloomfield}, which investigated the effect of patient self-testing and self-management of long-term anticoagulation on major clinical outcomes. Similarly, telemedicine services leverage IoT-enabled devices such as video conferencing tools and remote diagnostic equipment to facilitate virtual consultations between patients and healthcare providers, eliminating geographical barriers to care and improving access to medical expertise.

\item{ Improving Operational Efficiency and Resource Management}\\
IoT technology has the potential to transform hospital operations by improving efficiency, streamlining workflows, and optimizing resource utilization. Smart hospital infrastructure, powered by IoT devices and sensor networks, enables real-time monitoring and management of hospital facilities, equipment, and assets.
For example, IoT-enabled asset tracking systems can help hospitals efficiently manage medical equipment, track inventory levels, and prevent equipment loss. Similarly, smart building systems can optimize energy usage, improve facility maintenance, and enhance patient comfort and safety. By harnessing IoT technology, hospitals can achieve cost savings, operational efficiency gains, and a more sustainable healthcare infrastructure.

\item{Empowering Data-Driven Decision-Making} \\
In an era of big data and analytics, IoT technology plays a crucial role in enabling data-driven decision-making and predictive analytics in hospital settings. According to \cite{Raghupathi}, big data analytics in healthcare holds promise and potential for transforming healthcare delivery. By collecting and analyzing vast amounts of patient data in real-time, IoT devices provide healthcare providers with actionable insights into patient health trends, treatment efficacy, and resource allocation.

Predictive analytics algorithms, powered by IoT-generated data, can forecast patient outcomes, identify high-risk individuals, and prioritize interventions to prevent adverse events. For instance, predictive models can help hospitals anticipate patient readmissions, optimize bed management, and allocate resources more effectively. By leveraging IoT technology and analytics, hospitals can enhance clinical decision-making, improve patient care coordination, and ultimately, save lives.
\end{itemize}

\section{The IoT Integration Protocol for Enhanced Hospital Care}

The IoT Integration Protocol offers a structured approach to the integration of IoT technology into hospital workflows, encompassing a series of strategic steps and components. Research has highlighted various approaches and success stories in integrating IoT within hospitals to streamline operations and improve patient outcomes. For instance, implementing IoT protocols can significantly enhance patient care by providing real-time data for informed decision-making \cite{Akkaş}. Each step plays a crucial role in ensuring the successful deployment and utilization of IoT solutions within the healthcare environment. Let's delve deeper into each component:
\begin{enumerate}
    \item Assessment and Planning

The protocol begins with a comprehensive needs assessment and planning phase, where healthcare objectives, existing infrastructure, and use cases for IoT integration are identified and defined. A detailed implementation roadmap is developed to guide the deployment of IoT solutions within the hospital environment.
\item Device Selection and Integration

Next, appropriate IoT devices are selected based on defined use cases and objectives, ensuring interoperability and compatibility with existing systems. These devices are seamlessly integrated into hospital workflows, with a focus on data accuracy, reliability, and security.

\item Data Collection and Management

Once deployed, IoT devices collect and transmit patient data to centralized repositories, where robust data governance policies ensure the security, privacy, and integrity of patient information. Integration with electronic health records (EHR) systems enables seamless access to patient data for healthcare providers.

\item Analytics and Decision Support

The collected patient data is then analyzed using advanced analytics techniques to derive actionable insights and patterns. Clinical alerts and decision support tools are generated to assist healthcare providers in making informed decisions regarding patient care and treatment plans.

\item Continuous Monitoring and Evaluation

Throughout the implementation process, the protocol emphasizes continuous monitoring and evaluation of IoT-enabled systems, ensuring optimal performance and alignment with healthcare objectives. Clinical outcomes and operational metrics are regularly assessed to identify areas for improvement and optimization.

\item Training and Education

Finally, comprehensive training and education programs are provided to healthcare professionals and staff to promote the adoption and utilization of IoT technology. Collaboration among interdisciplinary teams fosters innovation and knowledge sharing, driving continuous improvement in care delivery and patient outcomes.
\end{enumerate}

The strategic steps outlined in the IoT Integration Protocol represent a systematic approach to leveraging IoT technology for enhanced patient care and operational efficiency within hospital settings. By meticulously planning and executing each phase, healthcare organizations can overcome challenges associated with technology adoption and maximize the benefits of IoT solutions.

One key aspect of the protocol is the emphasis on assessment and planning, which lays the groundwork for successful IoT integration. This phase enables healthcare organizations to align technology initiatives with strategic objectives and stakeholder needs, ensuring that IoT solutions address specific pain points and contribute to overarching goals. Furthermore, the comprehensive needs assessment allows for the identification of potential barriers to implementation, such as regulatory compliance, data security concerns, and interoperability issues. By proactively addressing these challenges during the planning phase, hospitals can mitigate risks and ensure a smoother transition to IoT-enabled workflows.

Another critical component of the protocol is continuous monitoring and evaluation, which underscores the importance of ongoing assessment and optimization. In a rapidly evolving healthcare landscape, the ability to adapt and refine IoT-enabled systems is essential for maintaining alignment with healthcare objectives and delivering value to stakeholders. Regular monitoring of clinical outcomes and operational metrics provides valuable insights into the effectiveness of IoT solutions, allowing hospitals to identify areas for improvement and optimization. Moreover, continuous evaluation enables healthcare organizations to stay abreast of technological advancements and best practices, ensuring that their IoT initiatives remain innovative and impactful in the long term.

In summary, the IoT Integration Protocol offers a structured framework for hospitals to harness the transformative potential of IoT technology. By following the strategic steps outlined in the protocol and incorporating analytical insights and continuous improvement efforts, healthcare organizations can drive innovation, improve patient outcomes, and enhance the overall quality of care delivery.
\section{Challenges and Strategies in Implementation
}

\subsection{Implementation Challenges}

While the IoT Integration Protocol offers a structured approach to IoT deployment, several challenges must be carefully navigated during implementation. These challenges represent critical considerations that healthcare organizations must address to ensure the successful integration of IoT technology into their workflows:
\begin{itemize}
    \item { Data Security and Privacy Concerns:}
    
    One of the foremost challenges is ensuring the security and privacy of patient data collected by IoT devices. Studies have identified multiple challenges such as data security, interoperability, and ethical considerations, emphasizing the need for robust strategies to address these issues \cite{Sadiku}. As hospitals adopt IoT solutions for patient monitoring and data collection, safeguarding sensitive health information becomes paramount. Healthcare organizations must implement robust security measures and encryption protocols to protect patient data from unauthorized access and cyber threats. In \cite{Mahmoud}, the current status, challenges, and prospective measures of IoT security are discussed, shedding light on the importance of mitigating security vulnerabilities to safeguard sensitive patient data and ensure the integrity and confidentiality of healthcare information. This underscores the critical need for healthcare organizations to prioritize security considerations when deploying IoT solutions.

    \item {Interoperability Issues:}
 
    Integrating IoT devices with existing IT systems poses significant challenges, including interoperability issues. As discussed in \cite{Tsai}  healthcare big data analytics, may hold complexities of seamless data exchange between IoT devices and legacy systems, due to the complex ecosystem of disparate systems and technologies. Overcoming interoperability issues requires careful planning, standardized protocols, and collaboration among vendors to ensure compatibility and integration with existing infrastructure. 
    
    \item {Ethical Considerations:}
 
     Ethical concerns surrounding patient consent, data ownership, and equitable access to healthcare services also merit attention. As IoT devices capture and analyze vast amounts of patient data, healthcare organizations must navigate ethical dilemmas related to data usage and sharing. Ensuring transparency, informed consent, and patient autonomy are essential principles that guide ethical decision-making in the context of IoT deployment.

    \item {Regulatory Compliance:} 
    
    Adhering to regulatory requirements and standards governing the use of IoT technology in healthcare settings presents a significant compliance challenge. Healthcare organizations must navigate a complex regulatory landscape, including HIPAA (Health Insurance Portability and Accountability Act) regulations and GDPR (General Data Protection Regulation) guidelines, to ensure compliance with data privacy and security regulations. Achieving regulatory compliance requires meticulous adherence to standards, policies, and best practices to mitigate legal risks and safeguard patient rights.
\end{itemize}

\subsection{Strategies for Overcoming Implementation Challenges}
In this section, we delve into analytical strategies and solutions tailored to address the multifaceted implementation challenges identified in Chapter 4.1. Recognizing the inherent complexities of healthcare IoT deployment, we explore nuanced approaches designed to navigate obstacles effectively. Key analytical considerations include:
\begin{itemize}

\item {Practical Cybersecurity Measures:}

An analytical approach to cybersecurity involves a comprehensive assessment of potential vulnerabilities and threats within healthcare IoT ecosystems. By conducting thorough risk assessments and threat modeling exercises, healthcare organizations can identify critical assets, assess their susceptibility to cyber threats, and prioritize security measures accordingly. Additionally, leveraging data analytics and machine learning techniques can enhance threat detection capabilities and enable proactive cybersecurity defenses.

\item {Interoperability Solutions:}

Analyzing interoperability challenges requires a nuanced understanding of data exchange protocols, system architectures, and stakeholder requirements. By employing analytical frameworks such as interoperability maturity models and data mapping techniques, healthcare organizations can identify interoperability gaps, assess compatibility issues, and develop strategic integration plans. Further-more, leveraging advanced analytics tools can facilitate real-time data harmoniza-tion and interoperability testing, ensuring seamless communication and data exchange among diverse IoT devices and systems.

\item {Ethical Frameworks:}

Ethical analysis in IoT deployment involves navigating complex ethical dilemmas surrounding patient consent, data privacy, and equity in healthcare access. By applying ethical decision-making frameworks such as principlism and conseque-ntialism, healthcare organizations can evaluate the ethical implications of IoT deployment and develop ethically sound policies and guidelines. Additionally, leveraging data analytics and privacy-enhancing technologies can enhance transpa-rency, accountability, and patient trust in healthcare IoT systems.

\item {Regulatory Compliance Strategies:}

Analyzing regulatory compliance challenges requires a systematic approach to understanding and interpreting healthcare regulations. By employing regulatory impact assessments and compliance gap analysis, healthcare organizations can assess the regulatory implications of IoT deployment and develop tailored compliance strategies. Furthermore, leveraging regulatory intelligence platforms and analytics tools can facilitate ongoing monitoring of regulatory changes and ensure timely adaptation to evolving compliance requirements.
\end{itemize}

\section{Impact Assessment and Effectiveness Evaluation}

\subsection{Impact Assessment}
Despite the previously-mentioned challenges challenges, the implementation of the IoT Integration Protocol holds the promise of significant benefits for hospitals and healthcare providers. Through rigorous impact assessment and evaluation, healthcare organizations can measure the tangible outcomes and value generated by IoT deployment:
\begin{itemize}
    \item {Improved Patient Outcomes:}

    IoT technology enables proactive monitoring and personalized interventions, allowing healthcare providers to identify health trends and predict adverse events. Real-time data insights lead to prompt interventions and better treatment outcomes, enhancing patient satisfaction.

    \item {Enhanced Operational Efficiency:}
    
    IoT integration streamlines workflows, optimizes resource utilization, and reduces administrative burden, leading to enhanced operational efficiency within healthcare organizations. Automated processes, remote monitoring capabilities, and predictive analytics enable hospitals to allocate resources more effectively, reduce wait times, and improve overall service delivery.

    \item {Increased Patient Satisfaction:}

    Through enhanced access to care, improved communication channels, and personalized treatment approaches, IoT deployment contributes to increased patient satisfaction. By empowering patients to actively participate in their healthcare journey and offering personalized care experiences, healthcare organizations can foster stronger patient-provider relationships and drive better health outcomes.

    \item{Cost Savings and Resource Optimization:}

    IoT-enabled care management strategies, reduced hospital admissions, and improved resource allocation contribute to cost savings and resource optimization within healthcare organizations. By leveraging IoT technology to optimize care delivery, hospitals can achieve greater operational efficiency, reduce healthcare costs, and improve the sustainability of healthcare delivery models.
\end{itemize}
In summary, while the implementation of the IoT Integration Protocol presents challenges, the potential benefits are significant. Through careful planning, strategic implementation, and continuous evaluation, healthcare organizations can harness the transformative power of IoT technology to improve patient outcomes, enhance operational efficiency, and drive innovation in healthcare delivery.

\subsection{Evaluating Impact and Effectiveness}

Moving forward, in this section we focus on analytical methodologies for evaluating the real-world impact and effectiveness of the IoT Integration Protocol. Drawing on advanced analytical techniques and data-driven insights, we assess various dimensions of impact, including:
\begin{itemize}
    \item {Quantitative Metrics:}
    
    Rigorous quantitative analysis enables us to measure the tangible benefits of the protocol on patient outcomes, operational efficiency, and cost savings. By employing statistical analysis and predictive modeling, we quantify the impact of IoT integration on key performance indicators such as patient readmission rates, length of stay, and resource utilization. Additionally, leveraging healthcare analytics platforms and data visualization tools allows for comprehensive trend analysis and benchmarking against industry standards.
    \item {Qualitative Insights:} 

    In-depth qualitative analysis provides valuable insights into stakeholder perspectives, experiences, and perceptions of the protocol's implementation. Through qualitative research methods such as interviews, focus groups, and thematic analysis, we capture nuanced insights into the protocol's perceived benefits, challenges, and areas for improvement from healthcare professionals, administrators, and patients. Additionally, leveraging sentiment analysis and natural language processing techniques can facilitate the extraction of actionable insights from unstructured data sources such as patient feedback and clinician narratives.

    \item{Case Studies and Success Stories:}

    Real-world case studies and success stories offer rich qualitative data to complement quantitative analysis and provide contextual understanding of the protocol's impact. By conducting comparative case studies and cross-sectoral analysis, we identify best practices, lessons learned, and success factors that contribute to the effective implementation of the protocol in diverse healthcare settings. Additionally, leveraging qualitative comparative analysis techniques allows for systematic comparison of case study findings and identification of common themes and patterns across different implementation contexts.

    \item{Long-Term Sustainability:}

    Analyzing the long-term sustainability of the protocol requires a holistic assessment of its scalability, adaptability, and resilience to changing healthcare dynamics. By employing sustainability impact assessments and scenario planning techniques, we evaluate the protocol's potential for long-term value creation and alignment with evolving healthcare needs and priorities. Additionally, leveraging predictive analytics and scenario modeling enables us to anticipate future trends and challenges, informing strategic decision-making and ensuring the protocol's continued relevance and effectiveness over time.
\end{itemize}

\section{Future Directions and Recommendations in Healthcare IoT Integration}

As the landscape of healthcare continues to evolve, the integration of Internet of Things (IoT) technology holds immense potential for transforming care delivery, improving patient outcomes, and enhancing operational efficiency. In this chapter, we explore future directions and provide recommendations to guide the advancement of healthcare IoT integration.

\begin{itemize}

\item Emerging Trends in Healthcare IoT \\
The rapid evolution of technology presents a multitude of opportunities and challenges for healthcare IoT integration. A comprehensive review highlights several potential future directions for IoT in healthcare, focusing on the integration of artificial intelligence (AI) and machine learning (ML). These technologies promise to enhance predictive analytics, facilitate early disease detection, and enable personalized treatment plans based on the vast amounts of data collected by IoT devices \cite{Chataut}. Ultimately, several emerging trends are poised to shape the future of healthcare delivery:
\begin{itemize}

\item AI and Machine Learning: \\ The integration of artificial intelligence (AI) and machine learning algorithms into IoT systems offers unprecedented capabilities for predictive analytics, clinical decision support, and personalized medicine. By harnessing the power of AI, healthcare organizations can unlock valuable insights from IoT-generated data and optimize care delivery processes.

\item Edge Computing: \\ Edge computing infrastructure brings computing resources closer to IoT devices, enabling real-time data processing, reduced latency, and enhanced data privacy. As healthcare IoT ecosystems become increasingly complex, edge computing solutions offer scalable and efficient architectures for managing and analyzing vast amounts of data at the point of care.

\item Blockchain Technology: \\ Blockchain technology holds promise for enhancing data security, integrity, and interoperability in healthcare IoT ecosystems. By providing a decentralized and immutable ledger for storing and sharing healthcare data, blockchain solutions offer transparency, traceability, and trust in data transactions, facilitating secure data exchange and collaboration among stakeholders.
\end{itemize}
\item Recommendations for Advancing Healthcare IoT Integration \\
To realize the full potential of healthcare IoT integration, concerted efforts are needed to address existing challenges and capitalize on emerging opportunities. The following recommendations provide actionable guidance for healthcare organizations, policymakers, and industry stakeholders:
\begin{itemize}

\item Investment in Interdisciplinary Research: \\ Encourage interdisciplinary research collaborations between healthcare professionals, engineers, data scientists, and policymakers to drive innovation and address complex challenges in healthcare IoT integration. By fostering cross-disciplinary partnerships, we can leverage diverse expertise and perspectives to develop innovative solutions that meet the evolving needs of healthcare delivery.

\item Promotion of Data Governance and Standards: \\ Establish robust data governance frameworks and interoperability standards to ensure the secure and seamless exchange of healthcare data across IoT devices and systems. By promoting standardized data formats, protocols, and interfaces, we can facilitate data interoperability, improve data quality, and enable interoperable communication among heterogeneous healthcare systems.

\item Incentivization of Healthcare IoT Adoption: \\ Provide incentives and support mechanisms to encourage healthcare organizations to adopt and implement IoT technology. This may include financial incentives, regulatory exemptions, and reimbursement models that reward value-based care, patient outcomes, and efficiency gains achieved through IoT integration. Additionally, fostering a culture of innovation and risk-taking within healthcare organizations can incentivize experimentation and adoption of new technologies.

\item Enhancement of Digital Literacy and Training: \\ Invest in digital literacy programs and training initiatives to equip healthcare professionals, administrators, and patients with the skills and knowledge needed to effectively leverage IoT technology. By providing comprehensive training on IoT devices, data analytics tools, and cybersecurity best practices, we can empower stakeholders to harness the full potential of healthcare IoT integration while mitigating potential risks and challenges.
\end{itemize}
\item Ethical Considerations and Policy Implications \\ 
 Ethical considerations and policy implications play a crucial role in shaping the future of healthcare IoT integration. Key considerations include:
\begin{itemize}
\item Privacy and Consent: \\ Ensure robust privacy protections and mechanisms for obtaining informed consent from patients regarding the collection, use, and sharing of their health data. By prioritizing patient privacy and autonomy, we can uphold ethical principles and build trust in healthcare IoT systems.

\item Equity and Access: \\ Address disparities in access to healthcare IoT technology and ensure equitable distribution of resources and benefits across diverse populations. By promoting inclusive design principles and considering the needs of marginalized communities, we can mitigate the risk of exacerbating existing healthcare inequalities.

\item Regulatory Frameworks: \\ Develop clear and comprehensive regulatory frameworks that balance innovation with patient safety, data privacy, and ethical considerations. By engaging stakeholders in the regulatory process and adapting regulations to the evolving landscape of healthcare IoT, we can create an enabling environment that fosters innovation while safeguarding patient rights and interests.
\end{itemize}
\item Collaboration and Knowledge Sharing \\
Collaboration and knowledge sharing are essential for driving progress and fostering innovation in healthcare IoT integration. Key initiatives include:
\begin{itemize}

\item Industry-Academia Collaboration: \\ Foster collaboration between industry partners and academic institutions to facilitate technology transfer, research collaboration, and talent development in healthcare IoT. By bridging the gap between academia and industry, we can accelerate the translation of research findings into real-world applications and promote the adoption of cutting-edge technologies in healthcare.

\item International Collaboration: \\ Promote international collaboration and knowledge exchange to address global challenges and leverage best practices in healthcare IoT integration. By sharing lessons learned, exchanging expertise, and collaborating on joint research initiatives, we can drive collective progress and amplify the impact of healthcare IoT on a global scale.

\item Community Engagement: \\ Engage patients, caregivers, and community stakeholders in the design, development, and implementation of healthcare IoT solutions. By fostering a culture of co-creation and user-centered design, we can ensure that healthcare IoT systems meet the diverse needs and preferences of end-users while promoting trust, transparency, and accountability in healthcare delivery.
\end{itemize}
\end{itemize}
\section{Conclusion}
In this concluding chapter, we embark on a comprehensive review of our exploration into healthcare IoT integration, dissecting its transformative potential and charting a course for the future.

Our journey through the realm of healthcare IoT integration has illuminated the profound impact this technology holds for revolutionizing patient care and operational efficiency within hospitals. Through the seamless integration of IoT devices, healthcare providers gain unprecedented access to real-time patient data, empowering them to deliver personalized care and make informed decisions that enhance patient outcomes. Simultaneously, IoT streamlines operational workflows, optimizing resource allocation and reducing administrative burdens, thus elevating the efficiency and effectiveness of healthcare delivery.

However, our exploration has not been without its challenges. Data security concerns, interoperability issues, and ethical considerations loom large on the path to IoT integration. Yet, by addressing these challenges head-on and fostering collaborative partnerships, we uncover opportunities for innovation and growth. By embracing emerging technologies like AI, edge computing, and blockchain, we can unlock new frontiers in healthcare delivery, while simultaneously ensuring that ethical principles and regulatory frameworks evolve in tandem with technological advancements.

As we look towards the future, it becomes clear that education and investment will be paramount in maximizing the potential of healthcare IoT integration. By equipping healthcare professionals with the necessary skills and knowledge to leverage IoT technology effectively, we can ensure that its benefits are fully realized. Additionally, continued investment in research and development will be essential in driving innovation and addressing the evolving needs of the healthcare landscape.

In conclusion, healthcare IoT integration represents a transformative force with the power to reshape the future of healthcare delivery. By navigating the challenges, embracing the opportunities, and remaining committed to the principles of responsible innovation and patient-centric care, we can unlock a future where healthcare is more efficient, more effective, and more equitable for all. As we embark on this journey, let us remain steadfast in our dedication to improving the lives of patients and communities worldwide, ensuring that the promise of healthcare IoT integration is realized to its fullest extent.


\begin{thebibliography}{99}
\markboth{}{}


\bibitem{Xiao}Li, Xiao, et al. "Digital health: tracking physiomes and activity using wearable biosensors reveals useful health-related information." PLoS biology 15.1 (2017): e2001402.

\bibitem{Alekya} Alekya, R., et al. "IoT based smart healthcare monitoring systems: A literature review." Eur. J. Mol. Clin. Med 7.11 (2021): 2020.

\bibitem{li2019digital} Li, Xiao, et al. "Digital health: tracking physiomes and activity using wearable biosensors reveals useful health-related information." PLoS biology 15.1 (2017): e2001402.



\bibitem{Bloomfield} Bloomfield, Hanna E., et al. "Meta-analysis: effect of patient self-testing and self-management of long-term anticoagulation on major clinical outcomes." Annals of internal medicine 154.7 (2011): 472-482.

 
\bibitem{Raghupathi} Raghupathi, Wullianallur, and Viju Raghupathi. "Big data analytics in healthcare: promise and potential." Health information science and systems 2 (2014): 1-10.

\bibitem{Akkaş} Akkaş, M. Alper, Radosveta Sokullu, and H. Ertürk Çetin. "Healthcare and patient monitoring using IoT." Internet of Things 11 (2020): 100173.

\bibitem{Sadiku} Sadiku, Matthew NO, Shumon Alam, and Sarhan M. Musa. "Internet of things in healthcare." IoT and ICT for Healthcare Applications (2020): 21-32.

\bibitem{Mahmoud} Mahmoud, Rwan, et al. "Internet of things (IoT) security: Current status, challenges and prospective measures." 2015 10th international conference for internet technology and secured transactions (ICITST). IEEE, 2015. 

\bibitem{Tsai} Tsai, Chun-Wei, et al. "Big data analytics: a survey." Journal of Big data 2.1 (2015): 1-32.

\bibitem{Chataut} Chataut, Robin, Alex Phoummalayvane, and Robert Akl. "Unleashing the power of IoT: A comprehensive review of IoT applications and future prospects in healthcare, agriculture, smart homes, smart cities, and industry 4.0." Sensors 23.16 (2023): 7194.




\end{thebibliography}
\end{document}